%% file: main.tex
\RequirePackage[table]{xcolor}

\documentclass[12pt]{article}
\usepackage{amssymb}   
\usepackage{amsmath}
\usepackage{amsthm}
\usepackage{graphicx}
\usepackage[sort]{cite}

\usepackage{multirow}
\usepackage{grffile}

\def\beq{\begin{equation}}   
\def\eeq{\end{equation}}
\def\bea{\begin{eqnarray}}  
\def\eea{\end{eqnarray}}

\def\f21{{}_2F_{1}}

\RequirePackage[colorlinks=true
,urlcolor=blue
,anchorcolor=blue
,citecolor=blue
,filecolor=blue
,linkcolor=blue
,menucolor=blue
,pagecolor=blue
,linktocpage=true
,pdfproducer=medialab
,pdfa=true
]{hyperref}

\newcommand{\cA}{\mathcal{A}}

\newcommand{\cL}{\mathcal{L}}
\newcommand{\cM}{\mathcal{M}}
\newcommand{\cE}{\mathcal{E}}

\newcommand{\cP}{\mathcal{P}}

\def\beq{\begin{equation}}
\def\eeq{\end{equation}}
\def\bsp#1\esp{\begin{split}#1\end{split}}

\newcommand{\ord}{{\mathcal{O}}}
\newcommand{\rd}{\textrm{d}}

\DeclareMathOperator{\E}{\textrm{E}}

\newtheorem*{thm}{Theorem}
\newtheorem*{corollary}{Corollary}

\usepackage[section]{placeins}

\usepackage{color}
\newcounter{RSQ}

\newlength{\parwidth}
\catcode`\@=11
\font\manfnt=manfnt
\def\Watchout{\@ifnextchar [{\W@tchout}{\W@tchout[1]}}
\def\W@tchout[#1]{{\manfnt\@tempcnta#1\relax%
  \@whilenum\@tempcnta>\z@\do{%
    \char"7F\hskip 0.3em\advance\@tempcnta\m@ne}}}
\let\foo\W@tchout
\def\dubious{\@ifnextchar[{\@dubious}{\@dubious[1]}}

\def\@dubious[#1]{%
  \setbox\@tempboxa\hbox{\@W@tchout#1}
  \@tempdima\wd\@tempboxa
  \list{}{\leftmargin\@tempdima}\item[\hbox to 0pt{\hss\@W@tchout#1}]}
\def\@W@tchout#1{\W@tchout[#1]}
\catcode`\@=12

\usepackage{longtable}

\begin{document}

\begin{flushright}
{\small
BONN-TH-2023-02

}
\end{flushright}

\vskip1cm
\begin{center}
{\Large \bf \boldmath Amplitude-like functions from entire functions}
\end{center}

  \vspace{0.5cm}
\begin{center}
\sc Claude~Duhr, \sc Chandrashekhar~Kshirsagar 
\\[6mm]
{\it Bethe Center for Theoretical Physics, Universit\"at Bonn, D-53115, Germany}
\\[0.3cm]
{\it Emails: cduhr@uni-bonn.de, chandra@uni-bonn.de}
\end{center}

\begin{abstract}
Recently a function was constructed that satisfies all known properties of a tree-level scattering of four massless scalars via the exchange of an infinite tower of particles with masses given by the non-trivial zeroes of the Riemann zeta function. A key ingredient in the construction is an even entire function whose only zeroes coincide with the non-trivial zeroes of the Riemann zeta function. In this paper we show that exactly the same conclusions can be drawn for an infinite class of even entire functions with only zeroes on the real line. This shows that the previous result does not seem to be connected to specific properties of the Riemann zeta function, but it applies more generally. As an application, we show that exactly the same conclusions can be drawn for $L$-functions other than the Riemann zeta function.
\end{abstract}

\newpage

\input{introduction.tex}

\input{acfs.tex}

\input{thm.tex}

\input{examples.tex}

\input{l-functions.tex}

\input{conclusions.tex}

\appendix
\input{app_S-class}

\section*{Acknowledgments}
The authors are greatful to Florian Loebbert for discussions. This work was co-funded by the European Union (ERC, LoCoMotive, 101043686). Views and opinions expressed are however those of the author(s) only and do not necessarily reflect those of the European Union or the European Research Council. Neither the European Union nor the granting authority can be held responsible for them.

\bibliography{bib}
\bibliographystyle{JHEP}

\end{document}

%% file: introduction.tex

\section{Introduction and motivation}
\label{sec:intro}

Scattering amplitudes are a cornerstone of modern research in Quantum Field Theory (QFT). First of all, they are one of the main ingredients to many computations for collider and gravitational wave phenomenology, and as such they play an important role when comparing theory and experiment. Second, already at tree-level they may provide important insight into the mathematical structure of QFTs. Sometimes they even serve as a tool to discover new theories with interesting properties and particle spectra, cf.,~e.g.,~ref.~\cite{Cheung:2014dqa}. Prominent examples of this are the Veneziano~\cite{Veneziano:1968yb} and Virasoro~\cite{Virasoro:1969me} amplitudes, which describe the the two-to-two scattering of open and closed strings at tree-level. These amplitudes feature an infinite number of poles, which represent the exchange of an infinite tower of (higher-spin) states with increasing masses.

Amplitudes are functions of the four-momenta of the scattering particles. In particular, for a two-to-two scattering of particles with mass $m_i$, the amplitude is a function of the usual Mandelstam invariants
\beq
s = (p_1+p_2)^2\,,\qquad t = (p_1+p_3)^2\,,\qquad u = (p_2+p_3)^2\,,
\eeq
constrained by 
\beq
s+t+u = \sum_{i=1}^4m_i^2\,.
\eeq
If the external states are scalars, the amplitude is a complex-valued function $\cM(s,u)$. The analytic structure of $\cM(s,u)$ is very much constrained from physics, and not every complex function in two variables may arise as a scattering amplitude in QFT or string theory. At tree-level, the function $\cM(s,u)$ must be meromorphic, with at most simple poles corresponding to the masses of the exchanged intermediate states.  Sometimes these constraints are powerful enough to uniquely determine amplitudes in specific theories. For example, the Veneziano amplitude is (loosely speaking) singled out as the unique amplitude with a prescribed high-energy behaviour and describing an infinite tower of higher-spin exchanges with unbounded mass spectrum~\cite{Caron-Huot:2016icg}. If some constraints are relaxed or changed, one may discover amplitudes in other theories. For example, if one relaxes the condition that the spectrum of exchanged high-spin states is unbounded, there is another solution, called the Coon amplitude~\cite{Coon:1969yw,Baker:1970vxk,Coon:1972qz}, which has received a lot of attention lately, cf.,~e.g.,~refs.~\cite{Figueroa:2022onw,Maldacena:2022ckr,Geiser:2022icl,Chakravarty:2022vrp,Cheung:2022mkw,Geiser:2022exp,Bhardwaj:2022lbz,Cheung:2023adk}.

It was recently proposed~\cite{Remmen:2021zmc} that it is possible to construct a function $\cM(s,u)$ consistent with known constraints on a two-to-two scattering of massless scalars, where the spectrum of exchanged particles is given by the non-trivial zeros of the Riemann zeta function. The latter is defined by
\beq\label{eq:zeta_def}
\zeta(z) = \sum_{n=1}^\infty\frac{1}{n^z}\,.
\eeq
This series converges absolutely for $\Re(z) > 1$. One can extend the definition by analytic continuation to $\mathbb{C}\setminus\{1\}$ via 
\beq\label{eq:zeta_cont}
\zeta(z) = 2^z\,\pi^{z-1}\,\sin(\pi z/2)\,\Gamma(1-z)\,\zeta(1-z)\,.
\eeq
One obtains in this way a meromorphic function with a simple pole at $z=1$, and holomorphic everywhere else. The Riemann zeta function possesses an infinite number of zeroes. First, it is easy to check that $\zeta(-2n)=0$ for every non-negative integer $n$. These are the so-called \emph{trivial zeroes}, and they are the only zeroes on the real line. In addition, there are zeroes in the complex plane (and they must come in complex-conjugate pairs). 
The celebrated \emph{Riemann hypothesis} expresses the remarkable conjecture that all non-trivial zeroes have the form $z_n = \frac{1}{2}\pm i\mu_n$, with $\mu_n$ real and positive. Note that, since $\zeta(z)$ is meromorphic, the set of zeroes must be discrete (and in particular the set of zeroes cannot have any accumulation point), and it is expected that there are infinitely many non-trivial zeroes. It was shown in ref.~\cite{Remmen:2021zmc} that, if we define
\beq\label{eq:cA_Remmen}
\cA(s) := -\frac{\rd}{\rd s}\log\Xi(\sqrt{s})\,,
\eeq
where $\Xi(z)$ is related to the Riemann zeta function via
\beq\label{eq:xi_def}
\Xi(z) := \xi\left({\frac{1}{2}}+iz\right)\,,\qquad \xi(z):=\frac{1}{2}z(z-1)\pi^{-z/2}\Gamma\left(\frac{z}{2}\right)\zeta(z)\,,
\eeq
then the function $\cM(s,t) := \cA(s)+\cA(u)$
has the properties of an amplitude describing the tree-level scattering of 4 massless scalars (we will review the complete set of constraints in section~\ref{sec:acfs}). This raises the intriguing question if there is a QFT for which $\cM(s,u)$ computes a scattering amplitude. 
Note that $\Xi(z)$ is even, $\Xi(-z)=\Xi(z)$, and entire, i.e., $\Xi(z)$ is holomorphic everywhere in the complex plane. It vanishes for $z =\pm\mu_n$, i.e., the zeroes of $\Xi$ are related to the non-trivial zeroes of the Riemann zeta function. This implies that $\cA(s)$ has simples poles at $s = \mu_n^2$. Hence, if there is a QFT for which $\cM(s,u)$ computes a scattering amplitude, then the spectrum of exchanged particles should be equal to the set of non-trivial zeroes of the Riemann zeta function (assuming the Riemann hypothesis holds, since otherwise the masses are complex)! This is in fact not the first time a connection was established between the Riemann hypothesis and the spectrum of a quantum mechanical system, cf.,~e.g.,~refs.~\cite{Montgomery,polya,berry,Srednicki:2011zz,Bender:2016wob,Sierra:2016rgn}.

The question if a QFT exists with mass spectrum $\mu_n$ and a four-point amplitude given by eq.~\eqref{eq:cA_Remmen} is a tough question, which is likely to remain open for a very long time. One may, however, try to answer a simpler question, namely whether one can find other complex functions $f$ from which one can build a function $\cM(s,u)$ that has all the properties of an amplitude in a QFT whose spectrum of exchanged particles is given by the set of zeroes of $f$. Finding such functions, or ruling out that they exist, may elucidate in how far the Riemann zeta function is special and how its properties are reflected in the putative QFT. In particular, ref.~\cite{Remmen:2021zmc} asked the question if it is possible to extend its construction to general Dirichlet $L$-functions (of which the Riemann zeta function is a special case), or even to arbitrary entire functions. However, no answer to these questions was given.

The goal of this paper is to give an answer to these questions. More precisely, we will show that, quite generically, we can associate to very large classes of even and entire functions $f$ with zeroes only on the real line a function $\cM_f(s,u)$ consistent with known constraints on the scattering of four (massless) scalars via the exchange of a spectrum of massive particles given by the zeroes of $f$. Our result allows us in particular to construct such a function $\cM_f$ for every Dirichlet $L$-function in the so-called Selberg class (assuming that the \emph{generalised Riemann hypothesis} holds for $L$-functions), which answers the questions asked in ref.~\cite{Remmen:2021zmc}. Our result shows in particular that the specific properties of the Riemann zeta function do not play an important role in the construction of $\cM_f(s,u)$, but they are a direct consequence of general properties of entire functions.

This paper is organised as follows: In section~\ref{sec:acfs} we review known constraints on two-to-two scattering amplitudes at tree-level, and we introduce the concept of \emph{amplitude-like} functions. Section~\ref{sec:thm} contains the main result of our paper: after some general review of entire functions and Hadamard's factorisation theorem in section~\ref{sec:entire_funcs}, we present our main theorem and its consequences in section~\ref{sec:thm_result}, and we prove our theorem in section~\ref{sec:proof}. In section~\ref{sec:examples} we illustrate our theorem on several examples, and in section~\ref{sec:L-functions} we use it to extend the results of  ref.~\cite{Remmen:2021zmc} from the Riemann zeta function to other types of $L$-functions. Finally, in section~\ref{sec:conclusions} we draw our conclusions.

%% file: acfs.tex

\section{Amplitude-like functions}
\label{sec:acfs}
The goal of this section is to review analytic properties of tree-level scattering amplitudes. We work in the setting of ref.~\cite{Remmen:2021zmc}, and we consider a two-to-two scattering of massless scalars described by an amplitude that can be cast in the form:
\beq\label{eq:fac_form}
\cM(s,u) = \cA(s)+\cA(u) = \cA(s)+\cA(-s-t)\,.   
\eeq
To be concrete, we may consider a theory with two distinct massless scalars $\phi_1$ and $\phi_2$, and we consider the scattering $\phi_1\,\phi_2\to\phi_1\,\phi_2$.\footnote{Note that not every theory of this type has necessarily amplitudes that can be cast in the form in eq.~\eqref{eq:fac_form}.}
In the following we discuss some general properties that any such amplitude must have:
\begin{enumerate} 
\item {\underline{Bose symmetry.}} For a scattering of the type $\phi_1\,\phi_2\to\phi_1\,\phi_2$, Bose symmetry implies $\cM(s,u) = \cM(u,s)$. This condition is automatically fulfilled if we work with the factorised form in eq.~\eqref{eq:fac_form}.
\item {\underline{Meromorphicity and simple poles on the positive real line.}}
 It is well known that tree-level amplitudes for scalar scattering are rational functions of the Lorentz invariant products of the four-momenta. 
 The only poles at finite values of the invariants arise from propagators going on shell, and locality dictates that all poles must be simple. In our scenario, this implies that $\cA(s)$ is a meromorphic function of $s$ with simple poles at most at $s=m_n^2\ge 0$. Note that the spectrum $m_n^2$ of exchanged particles cannot have any accumulation point, because a meromorphic function can only have isolated singularities.
\item {\underline{Negative residues at all poles.}} Close to the simple pole at $s=m_n^2$, the amplitude behaves like
$i\cM(s,u)\sim \frac{-ig^2}{s-m_n^2}$, where $g$ denotes the coupling constant of the interaction between the external scalars $\phi_1$, $\phi_2$ and the state $X_n$ of mass $m_n$ exchanged in the $s$-channel. If we want the coupling $\phi_1\phi_2X_n$ to be real, we must have $g^2>0$. Hence, we conclude that we must have
\beq
\textrm{Res}_{s=m_n^2} \cA(s) = \oint\frac{\rd s}{2\pi i}\cA(s) = -g^2 < 0\,,\textrm{~~~for all poles $s=m_n^2$.}
\eeq
\item {\underline{Polynomial boundedness.}} It is well known that in any QFT scattering amplitudes are polynomially bounded. More precisely, we must have for every fixed value of $t = -s-u$:
\beq
\lim_{s\to\infty}\cM(s,-s-t)s^{-N} = 0\,,\textrm{~~~for some positive integer $N$.}
\eeq
Here this is equivalent to 
\beq
\lim_{s\to\infty}\cA(s)s^{-N} = 0\,,\textrm{~~~for some positive integer $N$.}
\eeq
\item {\underline{Positivity constraints.}} In ref.~\cite{Adams:2006sv} it was shown that for $t=0$, we must have
\beq
\textrm{Res}_{s=0}\frac{\overline{\cM}(s,-s)}{s^{L+1}}=\oint\frac{\rd s}{2\pi i} s^{-N-1}\overline{\cM}(s,-s) > 0\,,
\eeq
where $\overline{\cM}(s,-s)$ is obtained from ${\cM}(s,-s)$ by subtracting poles at $s=0$.
By Bose symmetry, $\overline{\cM}(s,-s) = \overline{\cM}(-s,s) = \overline{\cA}(s)+\overline{\cA}(-s)$ is an even function (and $\overline{\cA}(s)$ is obtained by subtracting the poles at $s=0$), and so the previous equation only constrains the even Taylor coefficients of $\cA(s)$ around $s=0$. We can of course use the same argument to derive positivity constraints for $u=0$, which allows us to put constraints directly on $\cA(s)$:
\beq
\textrm{Res}_{s=0}\frac{\overline{\cM}(s,0)}{s^{L+1}}=\oint\frac{\rd s}{2\pi i} s^{-N-1}\overline{\cA}(s) > 0\,.
\eeq
It is easy to see this implies that all Taylor coefficients of  $\cA(s)$ around $s=0$ must be positive.
\end{enumerate}

We will refer to a function $\cM(s,u)$ that satisfies these five conditions as an \emph{amplitude-like function}.
In ref.~\cite{Remmen:2021zmc} it is shown that the function $\cM(s,u)$ constructed from the function $\cA(s)$ in eq.~\eqref{eq:cA_Remmen} satisfies these five constraints. As a consequence, $\cM(s,u)$
is amplitude-like.\footnote{In ref.~\cite{Remmen:2021zmc} only the positivity of the even Taylor coefficients was imposed. It is easy to see that  our requirement is stronger and strictly contains the positivity constraint on the even Taylor coefficients. As we will see later on, the function $\cA(s)$ from ref.~\cite{Remmen:2021zmc} (see eq.~\eqref{eq:cA_Remmen}) also satisfies our stronger requirement.}


%% file: thm.tex

\section{Amplitude-like functions from entire functions}
\label{sec:thm}

In this section we present the main result of our paper, namely we present a general construction of amplitude-like functions from a very large class of entire functions. We start by reviewing some mathematical background on entire functions and we present our result and discuss its consequences in section~\ref{sec:thm_result}. The proof is presented in section~\ref{sec:proof}.

\subsection{Entire functions}
\label{sec:entire_funcs}

In this section we review some standard material in complex analysis, in particular entire functions.
Recall that a function $f:\mathbb{C}\to \mathbb{C}$ is said to be \emph{entire} if it is holomorphic everywhere on the complex plane. Stereotypical examples of entire functions are polynomials and the exponential function. 

As a consequence of Liouville's theorem, every non-constant entire function must be unbounded. It will be useful to consider how an entire function behaves at infinity.
 We say that an entire function $f$ has \emph{order at most $\rho$} if there is $R>0$ and $C\ge0$ such that $|f(z)|<C\exp(|z|^\rho)$ for all $|z|>R$. The smallest such $\rho$ is called the \emph{order of $f$}. If $f$ has order at most $\rho$, then this means that $\log |f(z)|$ grows at most like $|z|^\rho$ for large $|z|$. This implies in particular that $\log|f(z)|$ is polynomially bounded:
 \beq
 \lim_{z\to \infty} z^{-d-1}\,\log|f(z)| = 0\,, 
 \eeq
 where we defined $d := \lfloor \rho\rfloor$, i.e., $d$ is the largest integer less or equal than $\rho$.
 In the following we will only consider entire functions of finite order.

An entire function may have zeroes, and since $f$ is holomorphic, its set of zeroes
cannot have any accumulation point. Every entire function of finite order $\rho$ can be cast in a standard form using \emph{Hadamard's factorisation theorem}:
\beq\label{eq:hadamard}
f(z) = e^{g(z)}\,z^m\,\prod_{n}\E_d\left(\frac{z}{z_n}\right)\,.
\eeq
where $d=\lfloor\rho\rfloor$, $g$ is a polynomial of degree at most $d$, $m$ is the order of $f$ at 0 and the product runs over all zeroes $z_n\neq0$ of $f$ counted with multiplicity. The function $\E_d(z)$ is the \emph{elementary factor}, defined by
\beq
\E_d(z) := \left\{\begin{array}{ll}
1-z\,, & \textrm{ if } d=0\,,\\
(1-z)\exp\left[\sum_{k=1}^d\frac{z^k}{k}\right]\,,&\textrm{ if } d>0\,.
\end{array}\right.
\eeq
Note that in the case where $f$ has an infinite number of zeroes, the product in eq.~\eqref{eq:hadamard} runs over an infinite number of terms. This case requires some careful consideration regarding the convergence of this infinite product. One can show (see, e.g.,~ref.~\cite{hadamard}) that the infinite product in eq.~\eqref{eq:hadamard} converges if and only if we have 
\beq\label{eq:convergence}
\sum_n\frac{1}{|z_n|^{d+1}}<\infty\,.
\eeq

In the following it will be useful to introduce the following notations.
Hadamard's factorisation theorem in eq.~\eqref{eq:hadamard} implies that we can write every function of finite order in the form
\beq
f(z) = z^m\,\cE_f(z)\,\cP_f(z)\,,
\eeq
where $\cE_f(z):= e^{g(z)}$ has no zeroes and $\cP_f(z) := \prod_{n}\E_d\left(\frac{z}{z_n}\right)$ has the form of an infinite product. 
Note that if $f_1$ and $f_2$ are two entire functions of order at most $\rho$, then so is their product, and we have
\beq
\cE_{f_1f_2}(z) = \cE_{f_1}(z) \cE_{f_2}(z) \textrm{~~~and~~~}\cP_{f_1f_2}(z) = \cP_{f_1}(z) \cP_{f_2}(z) \,.
\eeq
Finally, let us discuss some important consequence of eq.~\eqref{eq:convergence}. Consider a (possibly infinite) sequence $(z_n)_n$ of non-zero complex numbers such that eq.~\eqref{eq:convergence} holds. Then there is an entire function $f$ of order $d<\infty$ with precisely those zeroes. Indeed, Hadamard's theorem allows to easily construct such a function: it is simply the infinite product $\cP_f(z)$. In fact there are infinitely many such functions, and they differ precisely by an exponential factor $\cE_f(z)=e^{g(z)}$, where $g$ is a polynomial of degree at most $d$.

\subsection{The main result}
\label{sec:thm_result}
We now discuss our main result, which generalises the result of ref.~\cite{Remmen:2021zmc} from the entire function $\Xi$ in eq.~\eqref{eq:xi_def} to an infinite class of entire functions. We first need to restrict the class of entire functions $f$ that we will consider. First, the zeroes of $f$ will be related to the poles of the putative amplitude, so $f$ should only have zeroes on the real axis. Second, the propagator poles are related to  squared masses of the exchanged states, so we expect the zeroes to come in pairs $\pm z_n\neq0$. This gives, for functions of order at most $\rho$ (with $d=\lfloor \rho\rfloor$):
\beq\label{eq:E_d_sym}
\cP_{f}(z) = \prod_n\E_d\left(\frac{z}{z_n}\right)\E_d\left(-\frac{z}{z_n}\right) = \prod_n{\E}_{\lfloor d/2\rfloor}\left(\frac{z^2}{z_n^2}\right)\,,
\eeq
where in the last equality the product runs over the distinct zeroes of $f$ located on the positive real axis. Note that in this case $\cP_{f}(z)$ is an even function, $\cP_{f}(-z)=\cP_{f}(z)$, so that $\cP_{f}(\sqrt{z})$ defines an entire function of order at most $\rho/2$, with zeros of order $k_n$ at $z=z_n^2>0$. 

We will from now on focus on even entire functions $f(z)$ with zeroes on the real line. For such functions the zeroes always come in pairs $\pm z_n$, with $z_n>0$. In addition, $f$ may have a pole of order $m$ at $z=0$. Moreover, if $f$ is an even and entire function of order at most $\rho$, then $f(\sqrt{z})$ is en entire function of order at most $\rho/2$ (but it is not necessarily even). We then define (cf.~eq.~\eqref{eq:cA_Remmen}):
\beq
\cA_f(z) := -\frac{\rd}{\rd z}\log f(\sqrt{z}) = -\frac{\rd}{\rd z}\log\left[z^{m}\,\cE_f(\sqrt{z})\,\cP_{f}(\sqrt{z})\right]\,,
\eeq
where we use the notation 
\beq\label{eq:E_f_def}
\cE_f({z}) := \exp\left[\sum_{k=0}^{d} g_k\,z^k\right]\,,
\eeq
with $g_k$ some complex numbers. Note that, since $\cA_f(z) = \cA_{cf(z)}$ for every non-zero complex number $c$, we can assume without loss of generality $g_0=0$. 

The following notation will be useful:
\beq\label{eq:c_def}
c_{f,k} = \left\{\begin{array}{ll}
\sum_{n}\frac{1}{z_n^{2(k+1)}}\,, &\textrm{~~if~~}k \ge \lfloor d/2\rfloor\,,\\
0\,, &\textrm{~~if~~}k  <  \lfloor d/2\rfloor\,,
\end{array}\right.
\eeq
where $k$ is a positive integer.
It is easy to see that the series $\sum_{n}\frac{1}{z_n^{2(k+1)}}$ converges for $k\ge \lfloor d/2\rfloor$. Indeed, the only case that needs checking is when $f$ has an infinite number of zeroes $z_n$. Since the set of zeroes of $f$ has no accumulation point, the sequence $(|z_n|^2)_{n}$ is unbounded. Hence, there is some positive integer $N$ such that $|z_n|^2>1$ for $n>N$. This implies $|z_n|^{2(k+1)}\ge |z_n|^{2( \lfloor d/2\rfloor+1)}$, for all $n>N$ and $k\ge  \lfloor d/2\rfloor$. Convergence of $\sum_{n}\frac{1}{|z_n|^{2(k+1)}}$ then follows by comparing with the convergence criterion in eq.~\eqref{eq:convergence} applied to the infinite product in eq.~\eqref{eq:E_d_sym}.

Our main result is summarised in the following theorem:
\begin{thm} Let $f:\mathbb{C}\to\mathbb{C}$ be an even and entire function such that
\begin{enumerate}
\item $f$ has finite order $\rho$,
\item $f$ only has zeroes on the real line,
\item the $g_k$ are real and negative for all $1\le k\le \lfloor\rho\rfloor$.
\end{enumerate}
Then the function $\cM_f(s,u) := \cA_f(s)+\cA_f(u)$ is amplitude-like, i.e., it satisfies the 5 properties given in section~\ref{sec:acfs}.
\end{thm}
The proof of this theorem will be given in section~\ref{sec:proof} below. 
 Let us make some comments about the last condition, which is the most constraining one. First, we note that for an even function $f$, we have $g_{2k+1}=0$, so that the last condition really only applies to the even coefficients $g_{2k}$. Second, it is equivalent to $0\le \cE_f(z) < 1$ for all $z\in \mathbb{R}$. Finally, the last condition is always satisfied for even entire functions of order $\rho<2$. Indeed, if  $\rho<2$, we have $d\le 1$, and so eqs.~\eqref{eq:E_d_sym} and~\eqref{eq:E_f_def} imply:
\beq
\cP_f(z) = \prod_n\left(1-\frac{z^2}{z_n^2}\right)\textrm{~~~and~~~} \cE_f(z) = e^{g_0}=C\,,
\eeq
for some constant non-zero complex number $C$. Equivalently, we can write
\beq
f(z) = C\,z^{2m}\,\prod_n\left(1-\frac{z^2}{z_n^2}\right)\,.
\eeq
From here it is easy to see that the third condition of the theorem is always satisfied in this case, and we have:
\begin{corollary}
Let $f:\mathbb{C}\to\mathbb{C}$ be an even and entire function of finite order $\rho<2$ with zeroes only on the real line. Then the function $\cM_f(s,u)$ is amplitude-like.
\end{corollary}

In the remainder of this section we will discuss some general implications of our theorem.
First, we see can see that our theorem contains the results of ref.~\cite{Remmen:2021zmc} for the Riemann zeta function as a special case. Indeed, while the Riemann zeta function $\zeta(z)$ has a pole at $z=1$ (and is thus not entire), the function $\Xi(z)$ is an even and entire function of order 1, and we have:
\beq
\Xi(z) = \prod_{n=1}^{\infty}\left(1-\frac{z^2}{\mu_n^2}\right)\,,
\eeq
where the $\mu_n>0$ are the imaginary parts of the non-trivial zeroes of the Riemann zeta function. Assuming the Riemann hypothesis, the second condition of the theorem is satisfied.
Hence, the theorem applies, and $\cM_{\Xi}(s,u)$ is amplitude-like, in agreement with the findings of ref.~\cite{Remmen:2021zmc}.


We may ask if there are other entire functions that satisfy the assumptions of the theorem. In the following we argue that there are infinitely many such functions. Indeed, consider a sequence of positive real numbers $(z_n)_n$ without accumulation point such that $\sum_{n}\frac{1}{z_n^{d+1}}<\infty$. Then we know from Hadamard's factorisation theorem that there is an entire function $f$ of order at most $\rho$ with $d=\lfloor\rho\rfloor$ and with zeroes precisely at $z=\pm z_n$. We may pick:
\beq
f(z) = \cP_f(z) = \prod_n\E_d\left(\frac{z}{z_n}\right)\E_d\left(-\frac{z}{z_n}\right) = \prod_n{\E}_{\lfloor d/2\rfloor}\left(\frac{z^2}{z_n^2}\right)\,.
\eeq
It is easy to check that this function satisfies all the hypotheses of our theorem, and so the function $\cM_f(s,u)$ is amplitude-like. We thus see there is nothing special about the sequence $(\mu_n)_n$ of non-trivial zeroes of the Riemann $\zeta$ function, but the same conclusion holds for pretty much every sequence of real positive numbers that satisfy eq.~\eqref{eq:convergence}.
Finally, we mention that it is easy to see that if $f_1$ and $f_2$ satisfy the hypotheses of our theorem, then so does their product, and we have 
\beq\label{eq:amalgam}
\cA_{f_1f_2} = \cA_{f_1}+\cA_{f_2} \textrm{~~~and~~~} \cM_{f_1f_2} = \cM_{f_1}+\cM_{f_2}\,.
\eeq
Hence, also $\cM_{f_1f_2}$ 
 is amplitude-like.

\subsection{The proof of the theorem}
\label{sec:proof}
In this section we present the proof of our main theorem. We need to show that any $f$ that satisfies the hypotheses of the theorem also satisfies the five properties of section~\ref{sec:acfs}. Bose-symmetry (Property 1) is manifest, and there is nothing to check. 

Let $F:\mathbb{C}\to\mathbb{C}$ be a function analytic on some domain $U$. It is easy to see that
\beq
\frac{\rd}{\rd z}\log F(z) = \frac{F'(z)}{F(z)}
\eeq
has singularities at the zeros of $F$. If $F$ has a zero of order $M$ at $z=z_0\in U$, then $F'$ has a zero of order $M-1$ there, to that $F'/F$ has a simple zero at $z=z_0$. An easy application of the residue theorem shows that
\beq
\oint \frac{\rd z}{2\pi i}\,\frac{F'(z)}{F(z)} =  M > 0\,.
\eeq
We take $F=\cA_f$, and we have:
\beq
\cA_f(z) = -\frac{m}{z} - \frac{\cE_f'(z)}{\cE_f(z)} -\frac{\widetilde\cP_f'({z})}{\widetilde\cP_f({z})}\,,\qquad \widetilde\cP_f({z}):=\cP_f(\sqrt{z})\,.
\eeq
The first term clearly has a simple pole at $z=0$ with negative residue $-m$. The second term has no poles, because $\cE_f(z) = e^{g(z)}$ vanishes nowhere. The last terms has simple poles at $z=z_n^2$, with $z_n>0$. Hence,
\beq
\cA_f(z)  \sim \frac{-k_n}{z-z_n^2}\,,\qquad \textrm{for } z\sim z_n^2\,.
\eeq
We conclude that $\cA_f$ satisfies Properties 2 \& 3 of section~\ref{sec:acfs}.

Let us now check that $\cA_f$ is polynomially bounded (Property 4). Since $f$ has finite order $\rho$ say, we have $\cA_f(z) \sim (\rho-1)|z|^{\rho-1}$, and so $\cA_f$ is clearly bounded by $|z|^{\lfloor \rho\rfloor}$. In other words, the fact that $f$ has finite order immediately translates into $\cA_f$ being polynomially bounded.

It remains to show that $\cA_f$ satisfies Property 5. In other words, we need to show that all the Taylor coefficients of  $\overline{\cA}_f$ around $z=0$ are positive. The Taylor expansion of $\overline{\cA}_f$ is easy to obtain. Indeed, we have
\beq
\overline{\cA}_f(z) = \cA_f(z) + \frac{m}{z} = -\sum_{k=0}^{\lfloor d/2\rfloor-1}(k+1)g_{2k+2}\,z^{k}+\sum_{k=\lfloor d/2\rfloor}^\infty c_{f,k}\,z^{k}\,,
\eeq
We know that the infinite series defining $c_{f,k}$ in eq.~\eqref{eq:c_def} are all convergent and have positive summands, and so $c_{f,k}\ge 0$. Hence, the only non-trivial positivity constraints are those involving $g_{2k+2}$. Those are precisely satisfied if the hypotheses of the theorem hold. This finishes the proof.

%% file: examples.tex

\section{Examples}
\label{sec:examples}

In the previous section we presented a theorem which allows us to construct an infinite class of amplitude-like functions. Here we present several concrete examples, and we identify in each case a QFT for which $\cM_f(s,u)$ computes a scattering amplitude.

\subsection{Nowhere vanishing entire functions and contact interactions}
Consider a nowhere vanishing even and entire function $f$ of order $d$. By Hadamard's factorisation theorem, such a function is necessarily the exponential of a polynomial of order $d$:
\beq
f(z) = \cE_f(z) = \exp\left[\sum_{k=0}^{\lfloor d/2\rfloor}g_{2k}\,z^{2k}\right]\,.
\eeq
It is easy to see that such a function satisfies the hypotheses of the theorem, provided that $g_{k}\le0$, $1\le k\le d$ (recall that we may assume without loss of generality $g_0=0$). We have
\beq
\cA_f(z) = \sum_{k=0}^{\lfloor d/2\rfloor-1}(k+1)(-g_{2k+2})z^k\,.
\eeq
It is easy to check that the function
\beq
\cM_f(s,u) = \cA_f(s)  + \cA_f(u)  = \sum_{k=0}^{\lfloor d/2\rfloor-1}(k+1)(-g_{2k+2})\,\left(s^k+u^k\right)
\eeq 
computes the tree-level scattering amplitude $\phi_1\phi_2\to\phi_1\phi_2$ in the QFT described by the Lagrangian\footnote{We have computed the Feynman rules for small values of $d$ with FeynRules~\cite{Alloul:2013bka}.}
\beq
\cL = \frac{1}{2}\partial_{\mu}\phi_1\partial^{\mu}\phi_1+\frac{1}{2}\partial_{\mu}\phi_2\partial^{\mu}\phi_2 - \sum_{k=0}^{\lfloor d/2\rfloor-1}\lambda_k\,\phi_1\phi_2(\partial_{\mu_1}\cdots \partial_{\mu_k}\phi_1)(\partial^{\mu_1}\cdots \partial^{\mu_k}\phi_2)\,,
\eeq
with
\beq
\lambda_0 = -\frac{g_1}{4}\textrm{~~~and~~~} \lambda_k = (-1)^k\,2^{k-1}\,(k+1)\,(-g_{k+1})\,,\qquad >0\,.
\eeq
We see that nowhere vanishing entire functions that satisfy the hypotheses of our theorem describe a scattering of four scalars induced by contact interactions, and the dimensions of the contact operators are related to the order of the entire function. For example, the simplest non-constant such function is the exponential $f(z) = e^{-z^2}$ of order 2, and it describes the scattering of four scalars induced by the dimension-four operator $\phi_1^2\phi_2^2$. Another example would be $f(z) = e^{-z^4}$, which has order 4, and it describes a scattering induced by the dimension-six operator $\phi_1\phi_2\partial_\mu\phi_1\partial^{\mu}\phi_2$.


\subsection{Entire functions with zeroes}
Let us now turn to the case where $f$ has zeroes at $z=\pm z_n\neq 0$ (plus possibly a zero at the origin). It is sufficient to consider the case 
\beq
f(z) = z^m\cP_f(\sqrt{z}) = z^m\prod_n\E_{d'}\left(\frac{z}{z_n^2}\right)^{k_n}\,,
\eeq
where we defined $d':=\lfloor d/2\rfloor$. The product runs over the distinct zeroes $z_n\neq0$ of $f$ and $k_n$ denotes the multiplicity of that zero. Indeed, we know from the previous section that nowhere vanishing entire functions lead to contact interactions, and we can use eq.~\eqref{eq:amalgam} to construct the corresponding amplitude.

We start by noting that 
\beq
-\frac{\rd}{\rd z}\log\E_{d'}\left(\frac{z}{z_n^2}\right)^{k_n} =\sum_n \left(-\frac{k_n}{z_n^{2d'}}\right)\,\frac{z^{d'}}{z-z_n^2}\,,
\eeq
so that
\beq
\cA_f(z) = -\frac{m}{z}+\sum_n\left(-\frac{k_n}{z_n^{2d'}}\right)\,\frac{z^{d'}}{z-z_n^2}\,.
\eeq

Let us now discuss if there is a QFT with a scattering amplitudes for $\phi_1\phi_2\to \phi_1\phi_2$ given by $\cM_f(s,u) = \cA_f(s)+\cA_f(u)$. We will distinguish the two cases depending on the parity of $d'$.

If $d'=2\delta$ is even, it is straightforward to check that $\cM_f(s,u) = \cA_f(s)+\cA_f(u)$ computes the tree-level scattering $\phi_1\phi_2\to\phi_1\phi_2$ in the QFT described by the Lagrangian
\beq\bsp\label{eq:lag_even}
\cL = \frac{1}{2}\partial_{\mu}\phi_1\partial^{\mu}\phi_1 + \frac{1}{2}\partial_{\mu}\phi_2\partial^{\mu}\phi_2 + \sum_n&\frac{1}{2}\partial_{\mu}X_n\partial^{\mu}X_n - \frac{z_n^2}{2}\,X_n^2- \lambda_{\delta}^{\textrm{even}}\,\ord_{n,\delta}\,,
\esp\eeq
where we defined the operator
\beq\bsp
\ord_{n,\delta} &\,=  X_n\, (\partial_{\mu_1}\cdots\partial_{\mu_{\delta}}\phi_1)(\partial^{\mu_1}\cdots\partial^{\mu_{\delta}}\phi_2)\,,\qquad \lambda_{\delta}^{\textrm{even}} = \frac{2^{\delta}\sqrt{k_n}}{z_n^{2\delta}}\,.
\esp\eeq
We see that, just like in the case of nowhere vanishing entire functions, the order of $f$ is connected to the dimension of the operator. A special case is of course when $f$ is a polynomial.

If $d'$ is odd, then this does not work. Indeed, every Feynman diagram involves two insertions of the operator $\ord_{\delta}$, which necessarily leads to the even powers of $z$ in the numerator. One possibility is to consider instead a non-unitary version of the QFT described by eq.~\eqref{eq:lag_even}:
\beq\bsp
\cL = \frac{1}{2}\partial_{\mu}\phi_1\partial^{\mu}\phi_1 + \frac{1}{2}\partial_{\mu}\phi_2\partial^{\mu}\phi_2 + \sum_n&\partial_{\mu}X_n^{\dagger}\partial^{\mu}X_n - {z_n^2}\,X_n^{\dagger}X_n-X_n^{\dagger}\phi_1\phi_2- \lambda_{d'}^{\textrm{odd}}\ord_{n,d'}\,,
\esp\eeq
with
\beq
\lambda_{d'}^{\textrm{odd}} = \frac{2^{d'}k_n}{z_n^{2d'}}\,.
\eeq
Alternatively, one may consider a scattering of four distinct scalars, e.g., in a theory described by the Lagrangian
\beq
\cL = \sum_{i=1}^4  \frac{1}{2}\partial_{\mu}\phi_i\partial^{\mu}\phi_i + \sum_n\frac{1}{2}\partial_{\mu}X_n\partial^{\mu}X_n - \frac{z_n^2}{2}\,X_n^{2}-X_n\phi_3\phi_4- \lambda_{d'}^{\textrm{odd}}\ord_{n,d'}\,.
\eeq
It is then easy to check that the scattering amplitude for the process $\phi_1\phi_2\to\phi_3\phi_4$ is $\cM_f(s,u) = \cA_f(s)$.


%% file: l-functions.tex

\section{Amplitude-like functions from $L$-functions}
\label{sec:L-functions}

In section~\ref{sec:thm} we proved our main result, which allows us to construct amplitude-like functions from large classes of even and entire functions. This answers one of the questions asked at the end of ref.~\cite{Remmen:2021zmc}. Reference~\cite{Remmen:2021zmc} also asked the question if it was possible to extend its results from the Riemann zeta function to other (Dirichlet) $L$-functions. In this section we show that the answer to this question is positive and follows directly from our theorem.

There are various different classes of functions called $L$-functions in the mathematical literature. Many of these functions are defined axiomatically and belong to the so-called \emph{Selberg (S) class}. While we expect that our results related to amplitude-like functions remain true for all functions in the S-class, for simplicity of the exposition we restrict the discussion here to a subset of $L$-functions, the so-called Dirichlet $L$-functions, and we defer the discussion of the general case to appendix~\ref{app:S-class}.

A Dirichlet $L$-function can be defined through a series similar to the definition of the Riemann zeta function in eq.~\eqref{eq:zeta_def}:
\beq\label{eq:L-func}
L(z,\chi) = \sum_{n=1}^{\infty}\frac{\chi(n)}{n^z}\,,\qquad \Re(z)>1\,.
\eeq
Here $\chi(n)$ is a \emph{Dirichlet character}, i.e., a map $\chi: \mathbb{Z}\to \mathbb{C}$ satisfying the following conditions:
\begin{enumerate}
\item $\chi$ is multiplicative: $\chi(m\cdot n) = \chi(m)\chi(n)$\,,
\item $\chi$ is periodic with period $q$: $\chi(n+q) = \chi(n)$\,,
\item $\chi(n)$ is non zero only if $n$ and $q$ are co-prime, i.e., if gcd$(n,q)=1$.
\end{enumerate}
If the smallest period of $\chi$ is $q$, we say that $\chi$ is a character mod $q$. It is easy to check that every Dirichlet character mod $q$ must evaluate to a $q^{\textrm{th}}$ root of unity. Clearly, for the trivial character $\chi(n)=1$, $\forall n\in\mathbb{Z}$, the definition in eq.~\eqref{eq:L-func} reduces to the definition of the Riemann zeta function in eq.~\eqref{eq:zeta_def}. 

If $\chi_1$ is a character mod $q_1$, and if $q_1|q_2$, then we can define a character $\chi_2$ mod $q_2$ by $\chi_2(n) = \chi_1(n)$. The character $\chi_2$ is said to be \emph{induced by $\chi_1$}. A character that is not induced by any other character is called \emph{primitive}. In the following we restrict the discussion to primitive characters (because non-primitive characters are not expected to be in the $S$-class).

It is a fundamental property of $L$-functions that they satisfy functional equations similar to eq.~\eqref{eq:zeta_cont} for the Riemann zeta function.
If $\chi$ is a primitive Dirichlet character mod $q$, then the corresponding $L$-function satisfies the functional equation
\begin{equation}
\label{eq:L-func_eq}
L(z,\chi)=\frac{G(\chi)}{i^{\delta}}2^{z}\pi^{z-1}q^{-z}\sin\left(\frac{\pi}{2}(z+\delta) \right)\Gamma(1-z)L(1-z,\overline{\chi})\,,     
 \end{equation}
 where $\overline{\chi}$ is the complex conjugate of $\chi$, $\delta= \frac{1-\chi(-1)}{2}$, and
 $G(\chi)=\sum_{a=1}^{a=q}\chi(a)e^{2\pi i \frac{a}{q}}$ is the Gauss sum, which in case of primitive characters satisfies $|G(\chi)|=\sqrt{q}$. Just like in the case of the Riemann zeta function, the functional equation can be used to analytically continue the function to values $\Re(z)<1$. The $L$-function has zeroes in the complex plane. The \emph{generalised Riemann hypothesis} expresses the conjecture that the non-trivial zeroes of $L$ (i.e., those that are not captured by the prefactor in the functional equation~\eqref{eq:L-func_eq}) all lie on the critical line $\Re(z)=\frac{1}{2}$. 
 
Just like in the case of the Riemann zeta function, it is possible to construct a function that has zeroes only at the non-trivial zeroes of $L(z,\chi)$:
\begin{equation}
     \xi(z,\chi)=\left(\frac{q}{\pi}\right)^{\frac{z+\delta}{2}}\Gamma\left(\frac{z+\delta}{2}\right)L(z,\chi) \,.
 \end{equation}
 This function is analogous to the function $\xi(z)$ in eq.~\eqref{eq:xi_def}, and one can show that $ \xi(z,\chi)$ defines an entire function of order one (cf., e.g., ref.~\cite{S-class1}).
There is, however, an important difference between $L(z,\chi)$ for a non-trivial character $\chi$ and the Riemann zeta function. For the Riemann zeta function, the zeroes of $\xi(z)$ come in complex conjugate pairs $\frac{1}{2}\pm i\mu_n$, which is a necessary condition for the function $\Xi(z) := \xi\left(\frac{1}{2}+ iz\right)$ to be even. This in turn is one of the hypotheses for our theorem from section~\ref{sec:thm_result} to be applicable. This property does no longer hold for a non-trivial character $\chi$. Instead, if $\rho$ is a zero of $\xi(z,\chi)$, then so is $1-\bar{\rho}$ (and the generalised Riemann hypothesis implies $\rho = 1-\bar{\rho}$). As a consequence, the function $\xi\left(\frac{1}{2}+ iz,\chi\right)$ is in general not an even function of $z$, and so our theorem does not apply. Instead, we can consider the function
\beq
\Xi_{\chi}(z) := \xi\left(\frac{1}{2}+ iz,\chi\right)\,\xi\left(\frac{1}{2}+ iz,\overline{\chi}\right)\,.
\eeq
Since $\xi(z,\chi)$ and $\xi(z,\overline{\chi})$ are entire functions of order 1, the same holds true for $\Xi_{\chi}(z)$.
Using the functional equation~\eqref{eq:L-func_eq}, we can show that $\Xi_{\chi}(z)$ is an even function:
\beq
\Xi_{\chi}(-z) = \xi\left(\frac{1}{2}-iz,\chi\right)\,\xi\left(\frac{1}{2}- iz,\overline{\chi}\right)
= \xi\left(\frac{1}{2}+ iz,\overline{\chi}\right)\,\xi\left(\frac{1}{2}+ iz,{\chi}\right)= \Xi_{\chi}(z)\,,
\eeq
where the second step follows from the functional equation~\eqref{eq:L-func_eq}.
Hence, $\Xi_{\chi}(z)$ is an even entire function of order 1 with the Hadamard product representation
 \begin{equation}
    \Xi_{\chi}(z)=\Xi_{\chi}(0)\prod_{n} \left(1-\frac{z^2}{\mu_{\chi,n}^{2}}\right)^{k_{n}}\,,
\end{equation}
where $\frac{1}{2}+i\mu_{\chi,n}$ are the non-trivial zeroes of $L(z,\chi)$, and $k_n$ denote their multiplicity.
We can apply our theorem from section~\ref{sec:thm_result}, and we see that the following function is amplitude-like:
 \beq\bsp
\mathcal{A}_{\Xi_{\chi}}(s) &=  -\frac{\mathrm{d}}{\mathrm{d}s}\log{\Xi_{\chi}(\sqrt{s})} \\ 
 &= -\frac{i}{2\sqrt{s}}\Biggl[ \frac{L'(\frac{1}{2}+i\sqrt{s},\chi)}{L'(\frac{1}{2}+i\sqrt{s},\chi)}+ \frac{L'(\frac{1}{2}+i\sqrt{s},\overline{\chi})}{L'(\frac{1}{2}+i\sqrt{s},\overline{\chi})}  +\psi\left(\frac{1}{4}+\frac{i\sqrt{s}}{2}+\frac{\delta}{2}\right)+\log{\frac{q}{\pi}}         \Biggr]\\
&= \sum_{n}\frac{k_{n}}{-s+\mu_{\chi,n}^{2}}\,.
\esp\eeq
This shows that it is possible to construct an amplitude-like function not just for the Riemann zeta function, but for general $L$-functions, thereby answering the question asked in ref.~\cite{Remmen:2021zmc}. We emphasise that the functional equation~\eqref{eq:L-func_eq} plays an important role in proving that the function $\Xi_{\chi}$ is even. The discussion here strictly only applies tor Dirichlet $L$-functions with primitive character, which are special instances of the more general $L$-functions from the $S$-class. In appendix~\ref{app:S-class} we show that the arguments presented here can easily be extended to all $L$-functions from the $S$-class.

%% file: conclusions.tex

\section{Conclusions}
\label{sec:conclusions}

This paper was motivated by the recent ref.~\cite{Remmen:2021zmc}, where a function was constructed that has the properties of a tree-level scalar two-to-two scattering amplitude via the exchange of a tower of particles with a mass spectrum given by the non-trivial zeroes of the Riemann zeta function (assuming the Riemann hypothesis holds). Since the analytic structure of scattering amplitudes is constrained by physical considerations, the existence of such a function with the required properties seems (at first sight) surprising. This opens the intriguing possibility that there may be a QFT whose mass spectrum is related to the Riemann hypothesis. 
Reference~\cite{Remmen:2021zmc} did not provide any evidence for or against the existence of such a QFT, and answering this question will most likely remain beyond our abilities for a long time. Reference~\cite{Remmen:2021zmc} asked the question if it is possible to extend its construction to other classes of $L$-functions, or even to arbitrary entire functions. 

The purpose of our paper was to provide an answer to these questions. Our main result is a theorem which allows us to construct amplitude-like functions from very large classes of entire functions. The main tool is Hadamard's factorisation theorem, which allows one to represent every entire function $f$ of finite order as a product of an exponential factor and an infinite product that captures the location of the zeroes of $f$. Our theorem states that, under some mild assumptions on $f$, it is possible to construct an amplitude-like function $\cM_f$. The function-theoretic properties of $f$ directly translate into the constraints on scattering amplitudes from physics: the fact that $f$ has finite order implies the polynomial boundedness of $\cM_f$; the location of the zeroes determines the spectrum of exchanged particles; the positivity constraints on effective operators are related to the fact that Taylor coefficients can be represented as convergent series of positive powers. Finally, we showed that the exponential factor in the Hadamard factorisation is related to contact interactions, and the order of $f$ is related to the dimension of the operators that induce the scattering. 

Our theorem contains the result of ref.~\cite{Remmen:2021zmc} as a special case, and it immediately shows how to extend it to other classes of $L$-functions and entire functions.
This begs the question of what the implications are for the existence of a putative QFT with a scattering amplitude as given in ref.~\cite{Remmen:2021zmc}. While our theorem does not allow us to rule out the possibility that such a QFT exists, it implies that most likely there is nothing special about the non-trivial zeroes of the Riemann zeta function in this context, weakening arguments in favour of the existence of such a QFT. Another, and far more exciting possibility, could be that there are QFTs attached much more generally to entire functions, and the QFT attached to the Riemann zeta function is only a special case of a more general mechanism. Which of these two possibilities is correct goes beyond the scope of this paper, and could be the subject of future research.

%% file: app_S-class.tex
\section{Amplitude-like functions from the $S$-class}
\label{app:S-class}

In this appendix we show that the arguments of section~\ref{sec:L-functions} can be extended from Dirichlet $L$-functions for primitive characters to all $L$-functions from the $S$-class. In other words, we show that to every $L$-function from the $S$-class that satisfies the Riemann hypothesis we can associate an amplitude-like function. Before proving this statement, we start by briefly defining the $S$-class. 

\subsection{The Selberg class}

The {Selberg ($S$) class} are functions $F(z)$ that satisfy the following axions, called \emph{Selberg axioms} (see, e.g., refs.~\cite{Kaczorowski2006,S-class1}):
\begin{enumerate}
    \item \textbf{Dirichlet Series:} $F(z)$ can be written be written as an absolutely convergent Dirichlet series for $\Re(z)>1$:
 \begin{equation}
F(z)=\sum_{n=1}^{\infty}\frac{a(n)}{n^{z}}   \,. 
\end{equation}
   \item \textbf{Analytic Continuation:} There an integer $m\ge 0$ such that $(z-1)^{m}F(z) $ is an entire function of finite order.
   \item \textbf{Functional Equation:} $F$ satisfies the following functional equation:
   \beq \label{eq:Phi_func_eq}
\Phi(z)=\omega\, \overline{\Phi}(1-z) \,,
\eeq
where we defined $\overline{\Phi}(z) := \overline{\Phi(\bar{z})}$, and
   \begin{flalign}
\Phi(z)&=Q^{z}\prod_{j=1}^{r}\Gamma(\lambda_{j}z+\mu_{j})F(z)\,,
   \end{flalign}
with $\omega\in\mathbb{C}$, $|\omega|=1$, $\Re{\mu_{j}}\geq0$, $Q>0$, $\lambda_{j}>0$ and $r\ge0$.
   
   \item \textbf{Ramanujan hypothesis:} For every $\epsilon>0$, $a(n)\ll n^{\epsilon}$.
   \item \textbf{Euler Product:} For $\Re(z)>1$, 
  \begin{equation}
      \log{F(z)}=\sum_{n=1}^{\infty}\frac{b(n)}{n^{z}}\,,
  \end{equation}
  where $b(n)$ is non-zero only for $n=p^{l}$ where $p$ denotes a prime factor and $l\geq 1$, and $b(n)\ll n^{\theta}$ for $\theta<1/2$.
\end{enumerate}

The $\Gamma$ functions in eq.~\eqref{eq:Phi_func_eq} have poles, and so $F$ must have zeroes at the corresponding locations. These are the trivial zeroes of $F$.
Note that axiom 2 implies that $F$ has at most a pole of order $m$ at $z=1$, and it is holomorphic everywhere else. In the following we assume without loss of generality that $m$ is equal to the order of this pole. 
The statement of axiom 2 can actually be made even sharper. One can show that $(z-1)^mF(z)$ is an entire function of order 1~\cite{S-class1}. 

\subsubsection{Amplitude-like from the $S$-class}
Let now $F$ be a function from the $S$ class.
Let us define the following function:
\beq
\Theta(z) := z^m\,(1-z)^m\,\Phi(z)\,.
\eeq
Since $\Phi(z)\propto F(z)$, we can see that $\Theta$ has no pole at $z=1$, and therefore it is an entire function of order at most 1. 
It is also possible to show that $\Theta(z)$ has zeroes only at the non-trivial zeroes of $F$.
Moreover, it is easy to check that $\Theta$ satisfies the functional equation
\beq\label{eq:Theta_func_eq}
\Theta(z) = \omega\,\overline{\Theta}(1-z)\,.
\eeq
However, $\Theta$ will in general not be an even function. Instead, the functional equation~\eqref{eq:Theta_func_eq} implies
\beq
\Theta(z)\overline\Theta(z) = \overline\Theta(1-z)\Theta(1-z)\,.
\eeq
and therefore the function
\beq
\Xi_{F}(z) := \Theta\left(\frac{1}{2}+iz\right) \,\overline{\Theta}\left(\frac{1}{2}+iz\right)
\eeq
is even, $\Xi_{F}(-z) = \Xi_{F}(z)$. It then follows form the Corollary in section~\ref{sec:thm_result} that $\cM_{\Xi_F}(s,u)$ is amplitude-like, provided that the zeroes of $\Theta\left(\frac{1}{2}+iz\right)$ all lie on the real line, i.e., if $F$ satisfies the Riemann hypothesis.  Hence, we see that for all functions $F$ in the $S$-class satisfying these conditions, leads to an amplitude-like function $\cM_{\Xi_F}$.